**Speeds of sound for a biogas mixture CH$_4$+N$_2$+CO$_2$+CO from $p$ = (1 to 12) MPa at $T$ = (273, 300 and 325) K measured with a spherical resonator.**

Daniel Lozano-Martín[a], José J. Segovia[a,*], M. Carmen Martín[a], Teresa Fernández-Vicente[b], D. del Campo[b].

[a]TERMOCAL Research Group. University of Valladolid, Paseo del Cauce 59, 47011, Valladolid, Spain.

[b]Centro Español de Metrología, Alfar 2, 28760 Tres Cantos, Madrid, Spain.

* To whom correspondence should be addressed (e-mail): jose.segovia@eii.uva.es

**Abstract**

The present work aims to measure speeds of sound $c$ in a biogas mixture of CH$_4$+N$_2$+CO$_2$+CO, at $p$ = (1 to 12) MPa and $T$ = (273, 300 and 325) K, using a spherical acoustical resonator. The results are fitted to the virial acoustic equation of state, and the virial acoustic coefficients are obtained, $\beta_a$ and $\gamma_a$ and extrapolated to zero pressure, determining the adiabatic coefficient as perfect gas, $\gamma^{pg}$, and the isobaric and isochoric heat capacities as perfect gas, $C_p^{pg}$ and $C_V^{pg}$, respectively. The speeds of sound are acquired with a mean expanded relative uncertainty of 165 parts in 10$^6$ ($k$ = 2) and are compared with the results predicted by the reference equation of state for this kind of mixture (natural gas-like mixtures), EoS GERG-2008. Relative deviations between experimental data and values estimated by this model were less than 700 parts in 10$^6$ at $T$ = 325 K, and below 400 parts in 10$^6$, and within measurement uncertainty of at $T$ = 300 K, although appreciably higher at isotherm $T$ = 273 K at the highest pressure data for this work, and even reaching values above 3.400 parts in 10$^6$.

## 1. Introduction.

Within the framework of seeking to find energy sources to replace conventional fossil fuels, as well as making energy sources cheaper and with lower $CO_2$ emissions into the atmosphere and in addition to replacing depleted resources such as oil and coal in the long-term natural gas-like mixtures, such as biogas, have emerged as a possible alternative over the past few decades [1]. In order to further contribute to the implementation thereof, it is necessary to improve the thermodynamic models used to calculate and design the extraction, transportation, storage and distribution systems of biogas-like mixtures. The present research thus focuses on discussing how accurate the standard equation of state, EoS GERG-2008, proves to be when compared to measurements of the speed of sound through a biogas sample and with the thermodynamic properties (heat capacities) derived from these data, in addition to providing new accurate experimental results so that, should the scientific community consider it appropriate, new correlations may be made to improve the model.

In order to have a precise composition without impurities to achieve the lowest possible measurement uncertainty, a quaternary $CH_4+N_2+CO_2+CO$ mixture was chosen for the study rather than a gas sample obtained by a natural process of biomass degradation. Some works with other binary methane-like mixtures [2-4], have previously been reported by other authors.

The spherical resonator (operating at low resonance frequencies) has emerged as one of the best instruments to characterize the thermodynamic behavior of a fluid through speed of sound. This technique was developed by Moldover, Mehl and Greespan during the eighties [5], [6] and continued by Ewing and Trusler among others in the nineties [7], [8]. The resonator's acoustic model is described in section 2. The experimental set-up is shown in section 3 and the results obtained at three isotherms, 273, 300 and 325 K, and pressures range from 1 to 12 MPa are discussed in section 4 and compared with the GERG-2008 model.

## 2. Acoustic model.

The development of acoustic-physical techniques for the thermodynamic characterization of fluids is illustrated in full in [9]. The starting point is given by the equation:

$$c = \frac{2\pi a}{v_{0n}}(f_{0n} - \Delta f) \qquad (1)$$

where $a$ is the inner radius of the resonator cavity, $v_{0n}$ is the zero of the spherical Bessel first derivative of order 0 (these values are tabulated in [10]), $f_{0n}$ is the experimental resonance frequency of propagation mode $l$=0,$n$ (radial symmetric modes, the only non- degenerate) of the acoustic wave, and $\Delta f$ is the term modifying the model from the ideal case of perfect geometry and zero wall acoustic admittance (strictly, such corrections depend on the corrected frequency itself. Their behavior as corrector terms allows dependency on the frequency to be approximated to the experimental frequency $f_{0n}$). The contributions involved in $\Delta f$ taken into account in our study are summarized below, following the steps detailed in [6] and [9].

### 2.1. Thermal Boundary Layer Admittance, $y_{th}$.

This is due to the boundary conditions making the temperature and heat flux continuous in the interphase gas - cavity (in fact, because the thermal conductivity and heat capacity of the shell are much greater than for the biogas, the conditions mean that the temperature is constant throughout the wall). It causes a shift in the resonance frequency $\Delta f_{th}$ and a contribution to the half-width $g_{th}$, as [11]:

$$\frac{\Delta f_{th}}{f} = \frac{-(\gamma-1)}{2a}\delta_{th}\frac{1}{1-l(l-1)/v_{ln}^2} + \frac{(\gamma-1)}{a}l_{th} + \frac{(\gamma-1)}{2a}\delta_{th,w}\frac{\kappa}{\kappa_w} \qquad (2)$$

$$\frac{g_{th}}{f} = \frac{(\gamma-1)}{2a}\frac{1}{1-l(l-1)/v_{ln}^2}\delta_{th} + \frac{(\gamma-1)}{2a}\delta_{th,w}\frac{\kappa}{\kappa_w} \qquad (3)$$

being:

$l_{th}$, the thermal accommodation length:

$$l_{th} = \frac{\kappa}{p}\left(\frac{\pi MT}{2R}\right)^{1/2}\frac{2-h}{h}\frac{1}{C_v/R+1/2} \qquad (4)$$

where $h$ is the thermal accommodation coefficient. Although it depends on the nature of the gas and the resonator's material, $h$ is set equal to 1 and our results are assumed to be independent of its value, since our test of its influence shows negligible deviations of $c$ of less than 1 part in $10^7$ for different $h$ values. This is as expected since, compared to other authors who determined $h$ in argon [12] or in helium [13], we take measurements at 10 times higher pressures, where the contribution of $h$ should be about 10 times less important.

$\delta_{th}$, is the thermal penetration length in the fluid, which characterized the thickness of the thermal boundary layer:

$$\delta_{th} = \left(\frac{\kappa}{\pi \rho C_p f}\right)^{1/2} \tag{5}$$

$\delta_{th,w}$, is the thermal penetration length in the wall of the resonant cavity, which indicates the thickness of the thermal boundary layer in the shell (made of A321 stainless steel):

$$\delta_{th,w} = \left(\frac{\kappa_w}{\pi \rho_w C_{p,w} f}\right)^{1/2} \tag{6}$$

where $\gamma$ is the adiabatic coefficient, $\kappa$ and $\kappa_w$ are the thermal conductivity of the gas and the cavity wall, respectively; $M$ is the mean molar mass, $R$ is the gas constant, $C_v$ is the isochoric heat volume, $C_p$ and $C_{p,w}$ are the isobaric heat capacities of the gas and the shell wall, respectively, and $\rho$ and $\rho_w$ are the densities of the gas and the wall, respectively.

Since the wavelength $\lambda$ is much greater than the thermal penetration length at the frequencies of our work, the temperature gradients are larger in the thermal boundary layer than those in the fluid bulk such that the irreversible heat flows are also greater. The frequency shifts $\Delta f_{th}/f$ are between $-40$ parts in $10^6$ for the mode $(0,2)$ and $-20$ parts in $10^6$ for the mode $(0,6)$, decreasing slightly while the temperature increased at the same pressures.

## 2.2. Coupling of Fluid and Shell Motion Admittance, $y_{sh}$.

This takes into account the mechanical admittance of the cavity wall. Even if this term can be solved exactly for an isotropic spherical resonator, following the elastic behavior model of the resonator in [6] and [14], it was decided to use the simplified results shown in [9], because the system is allocated in high vacuum and we assume that acoustic radiation losses are negligible. Under this approximation, this correction term is a pure complex number, i.e., it only contributes with a frequency shift $\Delta f_{sh}$:

$$\Delta f_{sh} = -f \rho c^2 C / [1 - (f/f_{br})^2] \tag{7}$$

where $C$, the static compliance of the cavity and $f_{br}$, the breathing frequency, which is the lower radial symmetric mechanical resonance of the shell, are given by:

$$C = \frac{1-\sigma}{2[(b/a)^3 - 1]\rho_w c_w^2} \left(\frac{(b/a)^3}{1-2\sigma} + \frac{2}{1+\sigma}\right) \tag{8}$$

$$f_{br} = \frac{c_w}{2\pi a} \left( \frac{2[(b/a)^3 - 1]}{[(b/a) - 1][1 + 2(b/a)^3]} \right)^{1/2} \tag{9}$$

with $b$ being the outer radius of the cavity, $\sigma$ Poisson's ratio, and $c_w$ the sound speed of the wall material. The elastic properties of A321 steel have been approximated by those of grade A304 steel since more reliable data are found in the literature on the latter. The two grades are very similar (although A321 is stabilized with Ti) and their mechanical behavior is assumed to be equal within our work ranges. Density, Young's moduli and Poisson's ratio data are taken from [15], heat capacity and thermal conductivity from [16] and the thermal expansion coefficient from [17]. The frequency shift $\Delta f_{sh}/f$ takes values from $-200$ parts in $10^6$ for the mode (0,2) to $-600$ parts in $10^6$ for the mode (0,6) at high pressures, whereas ranges from $-20$ parts in $10^6$ for the mode (0,2) to $-60$ parts in $10^6$ for the (0,6) at the lower pressures. In sum, it is the main correction term on $f_{0n}$, since our studied biogas mixture is relatively dense at our range of measurements.

### 2.3. Duct Admittance, $y_0$.

Modification of cavity surface admittance by opening the biogas inlet feed duct ($y_0$) is modeled by a cylindrical tube of radius $r_0 = 0.5$ mm and length $L = 41$ mm, applying the Kirchhoff-Helmholtz theory for closed tubes finished in a rigid wall ($y_L = 0$), as described in [9]:

$$\Delta f_0 + ig_0 = \frac{c}{2\pi a} \frac{\Delta S}{4\pi a^2} iy_o \tag{10}$$

where:

$$y_0 = i\tan(k_{KH}L) \tag{11}$$

$$k_{KH} = \frac{\omega}{c} + (1 - i)\left(\frac{\omega}{2cr_0}[\delta_s + (\gamma - 1)\delta_{th}]\right) \tag{12}$$

with $\Delta S$ being the area of the opening, $\delta_s = \left(\frac{2D_s}{\omega}\right)^{1/2}$ and $\delta_{th} = \left(\frac{2D_{th}}{\omega}\right)^{1/2}$, where $D_s = \eta/\rho$ and $D_{th} = \kappa/\rho C_p$ are the shear viscous and thermal diffusivities, respectively; $\omega$ is the angular frequency and $\eta$ is the shear viscosity of the gas. As usual, the tube has been made of a length similar to the radius resonator, so as to minimize the effect of this perturbation (there are really two ducts in our experimental device: the inlet feed mentioned above and other blind tube which has no function). The frequency shifts $\Delta f_0 / f$ are around $-80$ parts in $10^6$ for the mode (0,2) and $-30$ parts in $10^6$ for the mode (0,6), increasing marginally when both frequency and temperature are increased and pressure is reduced.

### 2.4. Viscous Boundary Layer Admittance, $y_s$.

The admittance component due to the viscous boundary layer is zero for the radial symmetric modes ($l$=0) since the acoustic wave is normal incident on the wall and in the absence of tangential motion there is no viscous shear damping.

### 2.5. Dissipation in the Fluid.

The contribution to the half-width of the resonance peak caused by classical viscothermal mechanisms of acoustic energy absorption in the fluid bulk is given by:

$$g_{cl} = f^3 \frac{\pi^2}{c^2} \left[ \frac{4}{3} \delta_s^2 + (\gamma - 1) \delta_{th}^2 \right] \tag{13}$$

As already stated, energy dissipation due to heat flows has less effect in the fluid bulk than in the boundary layers because temperature and velocity gradients are lower in the fluid bulk and it is even less important at low frequencies since it is proportional to $\omega^2$, in contrast with the admittance component, due to the thermal boundary layer, which increases in proportion to $\omega^{1/2}$.

### 2.6. Geometrical Imperfections.

In addition to the non-zero surface shell admittance $y_s$, there are corrections over the ideal cavity by imperfect geometry, in the sense of perturbations of the perfect shell sphericity. As shown in [18], on the first-order perturbation theory the effect of a smooth spherical distortion does not produce a frequency shift of $f_{0n}$ for the radial modes if there is no change in volume. It is unnecessary to apply a higher order correction since the accuracy of parts in $10^6$ in the sound speed measurements is within the measurement uncertainty and can be reached with geometrical shape distortions of about $10\,\mu m$, which are feasible with the mechanical tolerances of conventional machining.

Nevertheless, the modification of the mechanical boundary impedance of the shell wall caused by the presence of the two transducers (source and detector, whose diaphragm motion is limited by its stiffness) has in fact been introduced in our computation, through frequency shifts $\Delta f_{tr}$ of the form shown in [11]:

$$\frac{\Delta f_{tr}}{f} = -\frac{\rho c^2 X_m r_{tr}^2}{2a^3} \tag{14}$$

where $r_{tr} = 1.5$ mm is the radius and $X_m = 7.1 \cdot 10^{-11}$ m/Pa is the compliance per unit area of the transducers. Even if $\Delta f_{tr}/f$ is a negligible contribution at low pressures ($\sim 5$ parts in $10^{\,6}$), it becomes the same order of magnitude as the others terms at the opposite case ($\sim 50$ parts in $10^6$).

### 2.7. Vibrational relaxation.

The last correction term, analyzed for the acoustic model used to describe our experimental data, is the contribution of molecular vibrational relaxation. To study its extent, we note that the excess half-widths $\Delta g$, defined as the difference between the experimental half-width $g$ and the contributions from the thermal boundary layer $g_{th}$, classical viscothermal absorption $g_{cl}$ and dissipation in the ducts $g_0$, are below the typical value of $10^{-5} \cdot f_{0n}$ of non-relaxing gases and do not show an increasing trend when pressure is decreased, as the whole measurement range illustrates in Figures 1-3 (with the exception of the mode (0,6), although this will not be taken into account in any case as will be seen later). Assuming that all the biogas molecules relax in unison with a single overall relaxation constant time $\tau_{vib}$ and that excess half-widths are entirely due to the vibrational effect, $\Delta g = \Delta g_{vib}$ can be written as:

$$\frac{\Delta g_{vib}}{f_{0n}} = \frac{(g-(g_{th}+g_{cl}+g_0))}{f_{0n}} = \frac{1}{2}(\gamma-1)\Delta\omega\tau_{vib} \tag{15}$$

where $\Delta = \sum_k x_k C_{vib,k}/C_p$ is the vibrational contribution to the isobaric heat capacity of the mixture and $C_{vib,k}$ is the molar isochoric vibrational heat capacity of each pure species $k$ of given composition $x_k$ which is estimated from a Planck-Einstein function with the vibrational frequencies reported in [19]. The relaxation constant times $\tau_{vib}$ are plotted in Figures 4-6, with mean values of $\bar{\tau}_{vib} = 2.28 \cdot 10^{-8}$ s at $T = 273$ K, $\bar{\tau}_{vib} = 2.47 \cdot 10^{-8}$ s at $T = 300$ K and $\bar{\tau}_{vib} = 2.10 \cdot 10^{-8}$ s at $T = 325$ K, four orders of magnitude lower than the resonance period $\tau \approx 1,8 \cdot 10^{-4} - 4,8 \cdot 10^{-5}$ s (with the exception mentioned above). The frequency dispersion of the resonance frequency $f_{0n}$ consequence of vibrational relaxation is therefore not introduced into the calculation.

### 2.8. Virial Acoustic Equation.

For the thermodynamic research of fluid properties, the speed of sound is related with the thermodynamic state by:

$$c^2 = \left(\frac{\partial p}{\partial \rho}\right)_S \qquad (16)$$

an expression which is only valid if no dispersion mechanism is taken into account, i.e., if $c(p,T)$ is frequency independent. This is true at sufficiently low frequencies and densities, on first approximation. In addition, $\omega \tau_c \ll 1$ is required where, for gas mixtures, the mean time between binary collisions, $\tau_c$, is close to the characteristic relaxation times of heat and momentum transfer, $\tau_{th} = D_{th}/c_0^2$ and $\tau_s = D_s/c_0^2$, respectively. In our biogas sample, $\tau_{th} \approx \tau_s \approx 10^{-12}$ s versus $\tau \approx 1.8 \cdot 10^{-4} - 4.8 \cdot 10^{-5}$ s, so the requirement is correctly met. Equation (16) can be expressed in terms of variables $(p,T)$ and $(\rho, T)$:

$$c^2 = \left[\left(\frac{\partial \rho}{\partial p}\right)_T - \frac{T}{\rho^2 c_p}\left(\frac{\partial \rho}{\partial T}\right)_p^2\right]^{-1} \qquad (17)$$

$$c^2 = \left[\left(\frac{\partial p}{\partial \rho}\right)_T - \frac{T}{\rho^2 c_v}\left(\frac{\partial p}{\partial T}\right)_\rho^2\right] \qquad (18)$$

The measurements are fitted to the acoustic virial equation of state, explicit in pressure or in density:

$$c^2 = A_0 + A_1 p + A_2 p^2 + \cdots \qquad (19)$$

$$c^2 = A_0(1 + \beta_a \rho + \gamma_a \rho^2 + \cdots) \qquad (20)$$

where the first acoustic parameter provides important information on the limit of perfect gas, $p \to 0$:

$$A_0 = \frac{RT\gamma^{pg}}{M} \qquad \rightarrow \qquad \frac{c_{p,m}^{pg}}{R} = \frac{\gamma^{pg}}{\gamma^{pg}-1} \qquad (21)$$

where the superscript "$pg$" indicates perfect-gas. As shown in [6], when proceeding to the acoustic virial equation regression, it is necessary to choose resonance modes of frequency that are far away enough from mechanical resonances (the breathing modes of the shell), so as to avoid high energy transfer from the acoustic wave through the cavity wall. If not the acoustic model developed is not valid. The frequency limit for our resonator is thus $f_{br} \approx 27500$ Hz, and so the research is confined to the first five radial modes of the speed of sound in the biogas.

### 3. Experimental equipment and process.

The spherical resonator used in this work is directly based on the designs developed by Trusler and Ewing [20] and employed recently in other research [21]-[22], where some meticulous schematic designs can be found. In order to avoid repetition, the apparatus is briefly described in these lines. The spherical resonant cavity is formed by the union of two aligned hemispheres, fixed by electron beam welding, fabricated from grade 321 austenitic stainless steel and manufactured at Imperial College, with nominal radius $a = 40$ mm and nominal thickness $b - a = 10$ mm. An average inner radius as a function of pressure and temperature was determined in a previous work [21] by calibrating with argon (of well-known equation of state). Although some authors have reported several problems with the junction of the hemispheres [23], because the welding does not usually fill the equatorial gap between hemispheres fully, this issue has been considered a negligible geometric imperfection as reported in [9].

Flush with the inner surface of the north shell hemisphere, two equal transducers of wide bandwidth, source and detector, are set at 90° between them to reduce overlapping between the radial mode (0,2) and the close degenerate mode (3,1). Following the design shown in [24], they are capacitance transducers of solid polymeric dielectric (a polyamide circular sheet) of $12 \, \mu m$ thickness and 3 mm diameter, coated with a gold layer of 50 nm thickness on the side facing the interior of the shell (the diaphragm). This is provided with a small perforation to allow the gas to pass to the interior volume of the transducer, thereby decreasing rigidity and increasing transducer sensitivity. To avoid problems, they are designed with mechanical resonance frequencies of the diaphragm around 40 kHz, well above the acoustic ones of the cavity.

The source transducer is driven by a pure alternate signal send out by the wave generator HP3225B, without bias voltage, so as to prevent direct electromagnetic coupling between the two transducers ("crosstalk") and producing an acoustic pressure response at twice the frequency of the wave synthesizer. It operates with root mean square voltage of 180 V on the source, delivered by a preamplifier because the wave generator only contributes with 40 V, with a precision in the frequencies of $1 \cdot 10^{-7} f$, as indicated by the manufacturer, although the synthesizer is plugged into a rubidium standard frequency, that improves it up to $5 \cdot 10^{-11} f$.

The transducer detector is fed with a bias voltage of 85 V whose exit signal is supplied to Lock-In SR850 DSP amplifier detector, after passing through a unity gain amplifier and high input impedance so as to remove the effect of the capacitance connection cables (triaxial cable with active guard), because the transducer capacitance $C_T$ is small enough (less than 100 pF) for a little load capacitance $C_L$, such as the connection cables, to imply a high division of the output signal by $(C_L + C_T)/C_L$. The Lock-In is referenced to the second harmonic of the driven source signal and measures amplitude $A$ and shift phase $\varphi$ of the detector signal, with amplitudes between $10\,\mu V - 200\,\mu V$, performed by acoustic pressure levels of $1\,mPa - 20\,mPa$, together with electronic noise levels of 50 nV rms. The signals in phase $u = A\cos\varphi$ and antiphase $v = A\sin\varphi$ are tracking at 11 equally spaced points in the interval $[f_N - g_N, f_N + g_N]$, in a ramp up and down. The total of 11 frequencies and 44 voltages are fitted to the equation:

$$u + iv = \sum_N \frac{if A_N}{(f^2 - F_N^2)} + B + C(f - f_0) \tag{22}$$

where $A_N, B$ and $C$ are complex constants and $N = 1$ for the radial symmetric modes and complex frequency values $F_N = f_N + ig_N$ are obtained with frequency deviations below $10^{-4} \cdot f_N$.

To control the thermal stability of the resonator, the resonant cavity is only linked by a copper block and is suspended in a cylindrical stainless steel vessel. This vessel is linked to the vacuum system, which uses a centrifugal Leybold Trivac B8B vacuum pump, in series with a turbomolecular pump Leybold SL300, reaching pressures about $10^{-3}$ Pa and avoiding convection heat transfer. Additionally, several aluminum layers over fiberglass with all possible wiring inside surrounds the jacket, thus minimizing radiation losses. The whole assembly is put into a Dewar with ethanol which is cooled by a Julabo FP89 thermal bath at $-30\,℃$, in an effort to transmit heat exclusively by thermal conduction through the copper block at the top. Temperature is established using three resistors: a band resistance of 10.6 Ω in the copper block in order to heat the resonant shell; a wire resistance of 28.5 Ω wound along the side of the jacket; and three band resistances of 174 Ω at the base of the jacket, in order to offset radiation losses. Temperature is controlled with three probes: an SPRT Rosemount 162D Pt-25 attached to the copper block and two SPRT Hart 5686 Pt-25 attached to the side and base of the jacket. The whole system is automated by three independent control loops implemented with PC Agilent VEE 7.0, which monitors the temperatures measured taken in the multimeter HP 3458A and the programmable HP E3632A power supplies plugged into the corresponding resistances. Usually, the whole apparatus takes one day to stabilize after each pressure drop, carried out in ramps of about 10 MPa, and about five days after each new assignment of reference temperature, ran at around 25 K ramps. With this thermostated device, a thermal gradient of only 1 mK between the two cavity hemispheres, with even lower drifts, is achieved during the measurement process.

Cavity pressure is gauged with a piezoelectric quartz transducer, Digiquartz 43RK-101, placed on the top of the biogas inlet tube (reported pressure data is corrected by the hydrostatic column), calibrated from 1 MPa to 20 MPa in TERMOCAL labs, this term being the main contribution to the pressure uncertainty which has relative expanded values of $\pm 2 \cdot 10^{-4}$ Pa/Pa ($k = 2$). The temperature cavity is measured with two SPRT Rousemont 162D of 25.5 Ω with Inconel X-750 encapsulated in a four-wire configuration, calibrated in our accredited laboratory, this term also being the main contribution to the temperature uncertainty which has expanded values of $\pm 5 \cdot 10^{-3}$ K ($k=2$). Each is attached at the top and bottom edges of the cavity to take into account the temperature gradient and they are plugged into an ac bridge ASL F18, referenced to external resistance Tinsley 5685A (of 25 Ω nominal, thermostated at 36 ºC and calibrated at INTA – Instituto Nacional de Técnica Aeroespacial).

In our spherical resonator, quality factors $Q = \frac{f_N}{2g_N} = resonance\ frequency/\ line-width$ are obtained with values between 1970.4 (mode (0,6) at $p = 1$ MPa, $T = 300$ K) and 14294.1 (mode (0,4) at $p$ = 13 MPa, $T$ = 300 K), typically one order of magnitude above resonators of other geometries, e.g. those of cylindrical cavity. They are also above other techniques such as interferometers, because at low frequencies these have large boundary effects which make accurate calculations difficult, while at high frequencies, they lose resolution since the resonance modes are not individually resolved (overlapping problems). The higher the $Q$, the sharper the resonance peaks, thus giving the best detection of the lines and resulting in greater accuracy of measurement.

The biogas mixture sample was synthesized at the CEM (Centro Español de Metrología) Reference Materials Laboratory by gravimetric method following the technical procedure described at standard EN ISO 6142:2006, whose composition is indicated in Table 1. Its theoretical critical points are: $T_c$ = 224.72 K, $p_c$ = 8.93 MPa and $\rho_c$ = 406.26 kg·m$^{-3}$.

## 4. Results and discussion.

The experimental results of the speed of sound are shown in Tables 2 to 4 together with the corresponding theoretical values determined using the equation of state (EoS) GERG-2008, now standard for biogas mixtures [25]-[26]. EoS GERG-2008, with the default settings of the reference fluid properties software NIST Refprop 9.1 (updated to the last version on June 10, 2014) [27], was used. The default values of the equation of state are not those of the original GERG model published in [26] for pure fluids, but a more complex and longer version, which is slightly more accurate. The expanded relative uncertainty in speed of sound is shown in Table 5, which takes into account seven contributions due to temperature and pressure of state-points, sample gas composition, resonant cavity radius calibration from speed of sound in Ar, resonance frequency fitting in biogas measures, mean standard deviation of sound speeds across modes in biogas sample and relative excess half-width from model. The latter term quantifies within the uncertainty in speed of sound our incomplete knowledge of the resonant cavity, i.e. how well the acoustical model shown in Section 2 describes the resonator acoustical behavior. Note that the combined expanded relative uncertainty ($k$ = 2) in speed of sound takes a value of 200 parts in 10$^6$, in line with that reported by other authors, and that radius calibration is the main contribution to these uncertainty. Not all the radial symmetric modes have been used to calculate the average value of $c(p,T)$ in each state, but the relative excess half-widths $\Delta g/f_{exp}$ plotted in Figures 1 to 3 are used as criteria. Thus, Figure 1 shows that the mode (0,5) must be discarded from the calculations at $T = 273$ K, because $\Delta g/f_{exp}$ is very high over the whole range of pressures. In addition, the mode (0,6) should be removed from calculations, because $\Delta g/f_{exp}$ does not tend to vanish with decreasing pressure. For the same reasons, modes (0,5) and (0,6) are also excluded at $T = 300$ K and mode (0,6) at $T = 325$ K, as shown in Figures 2 and 3.

In Table 6, the fitting parameters of the square speed of sound (corrected to specify them to the same reference isotherm) to the acoustic virial equation (equation 19) are shown. The standard deviation of the $n$ residuals of the fitted speed of sound in comparison to the experimental one, i.e., $\sigma = \sqrt{\Sigma\big((c_{\text{fitted}} - c_{\text{exp}})/c_{\text{exp}}\big)^2/(n-p)}$ , with $p$ number of parameters, were determined in order to choose the truncation order of the pressure series expansion, with the criteria that the corresponding $\sigma$ is within measurement uncertainty. The residuals between experimental and fitted values are shown in Figure 7 at each point. It is concluded that a fourth order acoustic model for $T = 273$ K and $T = 300$ K and a third order for $T = 325$ K are necessary. Including an additional term is negligible as shown in Table 7 for the different tested fittings. Table 6 also contains the expanded uncertainties of the parameters calculated through the Monte Carlo method, as indicated in [28].

Although not the main goal of this study, the thermodynamic properties derived from measurements of speed of sound in the limit to zero pressure, as perfect gas "pg", are computed and shown in Tables 8 to 10: the adiabatic coefficient $\gamma^{pg}$, the molar isochoric and isobaric heat capacities $C_V^{pg}$ and $C_p^{pg}$ and the second and third acoustic virial coefficients, $\beta_a$ and $\gamma_a$, respectively. All the results are compared and discussed with those reported by [27] using EoS GERG-2008 (standard for our fluid). A comparison with the AGA-8 and Peng-Robinson (PR) models is also provided in the tables to show the improvement of EoS-GERG against the other models predicting the acoustic virial coefficients. In any case, as expected it is satisfactory.

Their relative expanded uncertainties ($k$=2) are always better than 0.06 % for $\gamma^{pg}$, better than 0.26 % for $C_V^{pg}$ and $C_p^{pg}$, no worse than 1.6 % at 273 K, than 2.8 % at 300 K and than 2.4 % at 325 K for $\beta_a$ and between 10 - 40 % for $\gamma_a$. The uncertainty estimation of the fitting parameters grows very quickly together with the increasing order of the virial equation, from 1 part in 10⁴ to 1 part in 10, such that $u_r(A_2(T))$ significantly affects the uncertainty of $\gamma_a$.

Notice that the extrapolation to zero is performed on an adjustment about 1 to 10 MPa which may not seem a sufficiently low pressure to ensure good accuracy. However, experience from previous research, [21]-[22], using this technique with our experimental apparatus has shown that it is necessary to strike a balance between decreasing the pressure and measuring a strong enough signal to obtain the best accuracy. Below 1 MPa, the excess half-width begins to be too high, as can be seen in Figures 1 to 3, indicating that the assumptions of our acoustic model do not guarantee the quality of the results reported and, more importantly, the detector signal becomes so coarse that the measurement uncertainty significantly increases. Taking this into account, if we focus on the properties determined from $A_0(T)$, i.e., $\gamma^{pg}$, $C_V^{pg}$ and $C_p^{pg}$, the correlation to a fourth-order acoustic virial equation is more satisfactory at $T$ = 300 K than the fourth-order fitting at $T$ = 273 K or cubic fitting at $T$ = 325 K, because the relative deviation changes from 0.1 - 0.5 % to a relative difference with the theoretical values of 0.02 - 0.08 %, being within the measurement uncertainty in the latter case. Indeed, as the temperature increases the $\gamma^{pg}$ deviations also tend to increase passing from negative to positive values (respectively, $C_V^{pg}$ and $C_p^{pg}$ deviations tend to decrease). For $\beta_a$, relative deviations between the results calculated by any of the three equations of state and those measured are very similar and around 1 - 5 %, i.e., one order of magnitude above the measurement uncertainty but in agreement with other works [21]. However, the $\gamma_a$ calculation, performed by software Refprop 9.1, fails to determine its value using EoS GERG-2008, which seems to be an error of the calculation algorithm and, in any case, is not a problem with the model. In addition, the theoretical estimation of $\gamma_a$ given by EoS AGA-8 or Peng-Robinson (PR) is unsatisfactory, since deviations are about 100 % for AGA-8 and above 500 % for the cubic equation PR. For both acoustic virial coefficients, all the models overestimate the experimental data obtained in this work, no matter what the isotherm is.

In Figure 8, the relative deviations of measures $c(p,T)$ with respect to the theoretical calculations by EoS GERG-2008 are shown. The biggest differences are obtained at high pressure ($p = 12$ MPa) and low temperature ($T = 273$ K), where the results of model GERG-2008 overestimate in an order of magnitude of 1 part in $10^3$ of the research data. They are then reduced until they are cancelled at intermediate pressure and temperature ($p \sim 6$ MPa and $T = 300$ K). Then, they increase again at high temperature ($T = 325$ K) and low pressure ($p = 1$ MPa), where in this case the model GERG-2008 underestimates our measurements about 1 part in $10^4$. The orders of magnitude of the deviations from the EoS GERG-2008 are within the expected range. When the relative differences of this study are compared with those obtained in similar mixtures from reliable works, the overall average relative deviations, $\Delta_{average} = N^{-1} \sum_{i=1}^{N} (c_i - c_c)/c_i$ where $c_c$ is the speed of sound calculated from EoS-GERG and $N$ is the number of experimental points, are around 1 part in $10^4$ over a wide range of pressures and temperatures (with the exceptions and trends described below). For instance, Estela-Uribe et al. [3] obtained an average deviation $10^2 \cdot \Delta_{average} = 0.006$ at the sample $\{(1-x) \cdot CH_4 + x \cdot N_2\}$ with $x = 0.5422$, parallel to that established for the mixture $CH_4 + N_2 + CO + CO_2$ (Table 1) reported in these lines: $10^2 \cdot \Delta_{average} = 0.018$. Therefore, on the one hand, at the particular states of intermediate pressure and temperatures, the differences have a mean value $\sim 0.015$ % (similar to the overall $\Delta_{average}$) and are within measurement uncertainty. On the other hand, the remaining deviations are above measurement uncertainty but fall within the ranges of estimation predicted by the EoS-GERG2008, which are less than 0.1 % for temperatures ranging from (270 to 450) K at pressures up to 20 MPa and for temperatures from (250 to 270) K at pressures up to 12 MPa. Disagreement occurs at $T = 273$ K above 10 MPa, where $\Delta c = (c_{GERG} - c_{exp})/c_{exp}$ exceeds 0.3 %, far away from the limit estimated by GERG.

This behavior of the speed of sound is analogous to that described in [29] for the density in a similar mixture (synthetic coal methane mine mixture, CMM) performed with a single sinker densimeter with magnetic suspension in comparison with EoS GERG-2008. The relative deviation takes a maximum at the lowest temperatures (~250 K) and the highest pressures (~12 MPa), decreases and takes values close to zero at the middle pressure and temperature range and then increases again, underestimating the measurements at higher temperature (~375 K) and lower pressure (~1 MPa), as in our work. However, in [29] measurements at various pressures above those of our experiment were performed, the behavior was reversed. The relative differences between calculated and experimental data followed a new trend, decreasing for the lowest isotherms (250 K) and increasing for the highest temperatures (375 K) for the pressure range from 12 MPa up to 20 MPa. We assume that speed of sound at pressures higher than those in this research follows the same trend, since both thermodynamic properties, related by equations 17 or 18, were taken as the fundamental source of experimental data to model correlations of EoS GERG-2008 [27].

## 5. Conclusions.

The speed of sound through a quaternary synthesized biogas-like mixture of $CH_4+N_2+CO_2+CO$ was measured with expanded uncertainties better than $0.12 \ \mathrm{m \cdot s^{-1}}$ ($< 360 \ ppm$), within the order of magnitude of other works: [2], [3], [21] or [22]. These experimental data have been fitted to the acoustic virial equation.

Certain thermodynamic properties were then derived and compared with the calculations from the current reference equation of state for our mixture, GERG-2008 using RefProp 9.1: the adiabatic coefficient $\gamma^{pg}$ which deviates less than 0.1 %, the isochoric heat capacity as perfect gas $C_v^{pg}$ which differs less than 0.5 %, the isobaric heat capacity as perfect gas $C_p^{pg}$ which disagrees less than 0.4 % and the second acoustic virial coefficient $\beta_a$ which disagrees less than 5 %. The third acoustic virial coefficient $\gamma_a$ is discussed according to the calculations using EoS AGA-8 and EoS Peng-Robinson in addition to EoS GERG-2008, obtaining disparate results that do not adequately represent the experimental data in any condition.

The speed of sound behavior is closer to that modeled by EoS GERG-2008 at intermediate temperature and pressure ($p \sim 6$ MPa and $T = 300$ K), but smoothly tends to disagree at both low temperature and high pressure ($p \sim 12$ MPa and $T = 273$ K) and at the reverse state, high temperature and low pressure ($p \sim 1$ MPa and $T = 325$ K), with deviations up to 1 part in $10^3$. This is the same trend as described in [29] for the density of a similar mixture and it is expected that this difference begins to decrease with increasing pressure above our upper limit at 12 MPa, as in [29]. Although, measurements are within the limits of uncertainty stated by the EoS GERG-2008 in most cases, up to 0.1 % [26], this research aims to describe how well the model works and to provide accurate experimental data to improve it.

**Acknowledgements**

The authors wish to thank EURAMET and the European Union for supporting project ENG54, Metrology for Biogas, of the European Metrology Research Programme (EMRP) and to the Spanish Ministry of Economy and Competitiveness for the support to the project ENE2013-47812-R.

**Tables and Figures:**

**Table 1.** Molar composition of biogas sample and relative expanded uncertainties ($k$=2).

| Composition | $x_i \cdot 10^2$ / mol/mol | $U(x_i) \cdot 10^2$ / mol/mol |
|---|---|---|
| CO | 4.9899 | 0.0050 |
| $CO_2$ | 35.1484 | 0.0023 |
| $N_2$ | 10.0138 | 0.0040 |
| $CH_4$ | 49.8478 | 0.0069 |

**Table 2.** Experimental speeds of sound with their corresponding expanded uncertainties ($k$=2) and comparison with EoS GERG-2008 at $T$ = 273 K[(*)], for biogas with composition of Table 1, from the procedure described in Section 3 using the radial modes (0,2), (0,3) and (0,4), that are corrected using the acoustical model described in Section 2.

| $p$ / MPa | $T$ / K | $c_{exp}$ / m·s$^{-1}$ | $c_{GERG}$ / m·s$^{-1}$ | $10^6 \cdot (c_{GERG} - c_{exp})/c_{exp}$ |
|---|---|---|---|---|
| 11.586 | 272.969 | 313.359 | 314.440 | 3451.692 |
| 10.161 | 272.971 | 305.042 | 305.732 | 2262.929 |
| 9.127 | 272.966 | 301.941 | 302.475 | 1767.463 |
| 8.187 | 272.967 | 300.997 | 301.439 | 1466.261 |
| 7.155 | 272.960 | 301.643 | 302.005 | 1200.004 |
| 6.118 | 272.959 | 303.664 | 303.955 | 958.773 |
| 5.065 | 272.954 | 306.764 | 306.985 | 721.135 |
| 4.055 | 272.962 | 310.444 | 310.600 | 501.081 |
| 3.039 | 272.954 | 314.647 | 314.744 | 307.954 |
| 2.015 | 272.957 | 319.240 | 319.286 | 145.161 |
| 1.021 | 272.961 | 323.915 | 323.952 | 113.646 |

[(*)] Expanded uncertainties ($k$=2): $U(p) = 7.5 \cdot 10^{-5}$ ($p$/Pa) + 200 Pa; $U(T)$ = 4 mK; $U_r(c) = 2.0 \cdot 10^{-4}$ m·s$^{-1}$/ m·s$^{-1}$.

**Table 3.** Experimental speeds of sound with their corresponding expanded uncertainties ($k$=2) and comparison with EoS GERG-2008 at $T = 300$ K[(*)], for biogas with composition of Table 1, from the procedure described in Section 3 using the radial modes (0,2), (0,3) and (0,4), that are corrected using the acoustical model described in Section 2.

| $p$ / MPa | $T$ / K | $c_{exp}$ / m·s$^{-1}$ | $c_{GERG}$ / m·s$^{-1}$ | $10^6 \cdot (c_{GERG} - c_{exp})/c_{exp}$ |
|---|---|---|---|---|
| 12.826 | 299.852 | 339.69 | 339.77 | 229.92 |
| 12.074 | 299.846 | 336.039 | 336.091 | 157.054 |
| 11.030 | 299.845 | 332.10 | 332.08 | -62.27 |
| 10.161 | 299.849 | 329.708 | 329.685 | -70.637 |
| 9.104 | 299.842 | 327.872 | 327.851 | -65.787 |
| 8.079 | 299.848 | 327.116 | 327.095 | -65.199 |
| 7.061 | 299.838 | 327.249 | 327.221 | -87.432 |
| 6.062 | 299.850 | 328.121 | 328.077 | -134.539 |
| 5.050 | 299.841 | 329.638 | 329.570 | -206.866 |
| 4.029 | 299.850 | 331.715 | 331.615 | -300.332 |
| 3.020 | 299.850 | 334.211 | 334.077 | -401.895 |
| 2.010 | 299.851 | 337.063 | 336.906 | -465.456 |
| 1.004 | 299.845 | 340.187 | 340.035 | -444.788 |

[(*)] Expanded uncertainties ($k$=2): $U(p) = 7.5 \cdot 10^{-5}$ ($p$/Pa) + 200 Pa; $U(T) = 4$ mK; $U_r(c) = 2.0 \cdot 10^{-4}$ m·s$^{-1}$/ m·s$^{-1}$.

**Table 4.** Experimental speeds of sound with their corresponding expanded uncertainties ($k$=2) and comparison with EoS GERG-2008 at $T$ = 325 K[*],for biogas with composition of Table 1, from the procedure described in Section 3 using the radial modes (0,2), (0,3), (0,4) and (0,5), that are corrected using the acoustical model described in Section 2.

| $p$ / MPa | $T$ / K | $c_{exp}$ / m·s$^{-1}$ | $c_{GERG}$ / m·s$^{-1}$ | $10^6 \cdot (c_{GERG} - c_{exp})/c_{exp}$ |
|---|---|---|---|---|
| 11.094 | 324.749 | 351.323 | 351.244 | -222.384 |
| 10.035 | 324.740 | 349.125 | 348.989 | -389.304 |
| 9.082 | 324.743 | 347.714 | 347.584 | -374.234 |
| 8.094 | 324.741 | 346.843 | 346.712 | -378.400 |
| 7.055 | 324.744 | 346.522 | 346.385 | -395.874 |
| 6.056 | 324.744 | 346.743 | 346.589 | -443.393 |
| 5.043 | 324.744 | 347.438 | 347.264 | -501.457 |
| 4.045 | 324.744 | 348.54 | 348.34 | -581.73 |
| 3.031 | 324.743 | 350.034 | 349.808 | -645.645 |
| 2.038 | 324.748 | 351.820 | 351.573 | -702.041 |
| 1.011 | 324.742 | 353.96 | 353.70 | -719.85 |

[*] Expanded uncertainties ($k$=2): $U(p)$ = 7.5·10$^{-5}$ ($p$/Pa) + 200 Pa; $U(T)$ = 4 mK; $U_r(c)$ = 2.0·10$^{-4}$ m·s$^{-1}$/ m·s$^{-1}$.

**Table 5.** Uncertainty budget for the speed of sound measurements. Unless otherwise specified, uncertainties are indicated with a coverage factor $k$=1.

| Source | | Magnitude | Contribution to speed of sound uncertainty, $10^6 \cdot u_r(c)$ / (m·s$^{-1}$)/( m·s$^{-1}$) |
|---|---|---|---|
| | | *State-point uncertainties* | |
| Temperature | Calibration | ±0.002 K | |
| | Resolution | ±7.2·10$^{-7}$ K | |
| | Repeatability | ±5.6·10$^{-5}$ K | |
| | Gradient (across hemispheres) | ±3.3·10$^{-4}$ K | |
| | Sum | ±0.002 K | ±4.6 |
| Pressure | Calibration | ±(7.5·10$^{-5}$·p + 2·10$^{-4}$) MPa | |
| | Resolution | ±2.9·10$^{-5}$ MPa | |
| | Repeatability | ±4.0·10$^{-5}$ MPa | |
| | Sum | ±(5.9·10$^{-4}$ to 1.5·10$^{-4}$) MPa | ±2.6 |
| Gas composition | Purity | ±1.2·10$^{-6}$ kg/mol | |
| | Molar mass | ±7.3·10$^{-7}$ kg/mol | |
| | Sum | ±1.4·10$^{-6}$ kg/mol | ±24.9 |
| | | *Cavity radius* | |
| Radius from speed of sound in Ar | Frequency fitting | ±3.3·10$^{-6}$ m | |
| | Dispersion of modes | ±7.9·10$^{-7}$ m | |
| | Speed of sound from bibliography | ±4.0·10$^{-6}$ m | |
| | Sum | ±5.5·10$^{-6}$ m | ±76.3 |
| | | *Fitting of, and corrections to, resonance frequency* | |
| Frequency fitting | | ±0.97 Hz | ±53.6 |
| Dispersion of modes | | ±4.5·10$^{-3}$ m·s$^{-1}$ | ±13.7 |
| Relative excess half-width | | ±3.0·10$^{-5}$ Hz/Hz | ±16.4 |
| Sum of all contributions to $c$ | | | ±99 |
| $10^6 \cdot U_r(c)$ / (m·s$^{-1}$)/( m·s$^{-1}$) [(*)] | | | ±200 |

[(*)] With coverage factor $k$=2.

**Table 6.** Fitting parameters of square speed of sound according equation (19) and their corresponding expanded uncertainties ($k$=2).

| $T$ / K | $A_0(T)$ / m²·s⁻² | $A_1(T)$ / m²·s⁻²·Pa⁻¹ | $A_2(T)$ / m²·s⁻²·Pa⁻² | $A_3(T)$ / m²·s⁻²·Pa⁻³ | $A_4(T)$ / m²·s⁻²·Pa⁻⁴ |
|---|---|---|---|---|---|
| 273 | 108147±67 | (-323.3±5.3)·10⁻⁵ | (6.8±1.4)·10⁻¹¹ | (-1.3±1.5)·10⁻¹⁸ | (113.1±5.4)·10⁻²⁶ |
| 300 | 117961±71 | (-228.4±6.2)·10⁻⁵ | (4.6±1.8)·10⁻¹¹ | (7.6±2.0)·10⁻¹⁸ | (11.4±7.6)·10⁻²⁶ |
| 325 | 126937±66 | (-170.8±4.2)·10⁻⁵ | (67.4±7.4)·10⁻¹² | (52.3±3.9)·10⁻¹⁹ | - |

**Table 7.** Standard deviations ($\sigma$) of the residuals of the different tested fittings to the acoustic virial equation.

| | $\sigma\ [(c_{\text{fitted}} - c_{\text{exp}})/c_{\text{exp}}]$ / parts in 10⁶ | | | |
|---|---|---|---|---|
| $T$ / K | 3-parameters without $p^{-1}$ [a] | 4-parameters [b] | 3-parameters with $p^{-1}$ [c] | 5-parameters [d] |
| 273 | 4248.0 | 470.8 | 2634.8 | 54.4 |
| 300 | 1913.8 | 70.7 | 1131.4 | 34.8 |
| 325 | 546.2 | 22.0 | 290.9 | 21.6 |

[a] $c^2(p,T) = A_2(T)\cdot p^2 + A_1(T)\cdot p + A_0(T)$; [b] $c^2(p,T) = A_3(T)\cdot p^3 + A_2(T)\cdot p^2 + A_1(T)\cdot p + A_0(T)$;

[c] $c^2(p,T) = A_2(T)\cdot p^2 + A_1(T)\cdot p + A_0(T) + b(T)\cdot p^{-1}$; [d] $c^2(p,T) = A_4(T)\cdot p^4 + A_3(T)\cdot p^3 + A_2(T)\cdot p^2 + A_1(T)\cdot p + A_0(T)$.

**Table 8.** Thermodynamic derived properties from the first acoustic virial coefficient (equations 20 and 21) at $T$ = 273 K with their corresponding expanded uncertainties ($k$=2) and comparison with the different EoS.

| $T$ = 273 K | $Z_{exp}$ | $(Z_{exp}- Z_{GERG})/Z_{exp}$ % | $(Z_{exp}-Z_{AGA})/Z_{exp}$ % | $(Z_{exp}- Z_{PR})/Z_{exp}$ % |
|---|---|---|---|---|
| $\gamma^{pg}$ | 1.31826±0.00085 | -0.05 | -0.05 | -0.05 |
| $C_v^{pg}$ / $J\cdot mol^{-1}\cdot K^{-1}$ | 26.125±0.070 | 0.19 | 0.21 | 0.19 |
| $C_p^{pg}$ / $Jmol^{-1}\cdot K^{-1}$ | 34.439±0.094 | 0.15 | 0.16 | 0.15 |
| $\beta_a$ / $m3\cdot mol^{-1}$ | $(-67.9\pm1.1)\cdot10^{-6}$ | -0.85 | -5.5 | -8.5 |
| $\gamma_a$ / $(m3\cdot mol^{-1})^2$ | $(32.4\pm6.6)\cdot10^{-10}$ | $-4.3\cdot10^9$ | -130 | -470 |

**Table 9.** Thermodynamic derived properties from the first acoustic virial coefficient (equation 20 and 21) at $T$ = 300 K with their corresponding expanded uncertainties ($k$=2) and comparison with the different EoS.

| $T$ = 300 K | $Z_{exp}$ | $(Z_{exp}- Z_{GERG})/Z_{exp}$ % | $(Z_{exp}-Z_{AGA})/Z_{exp}$ % | $(Z_{exp}- Z_{PR})/Z_{exp}$ % |
|---|---|---|---|---|
| $\gamma^{pg}$ | 1.30848±0.00082 | 0.02 | 0.014 | 0.02 |
| $C_v^{pg}$ / $J\cdot mol^{-}1\cdot K^{-1}$ | 26.953±0.071 | -0.08 | -0.06 | -0.08 |
| $C_p^{pg}$ / $J\cdot mol^{-}1\cdot K^{-1}$ | 35.267±0.096 | -0.06 | -0.05 | -0.06 |
| $\beta_a$ / $m^3\cdot mol^{-1}$ | $(-48.3\pm1.3)\cdot10^{-6}$ | -5.0 | -10 | -13 |
| $\gamma_a$ / $(m3\cdot mol^{-1})^2$ | $(24.3\pm9.5)\cdot10^{-10}$ | $-4.0\cdot10^9$ | -180 | -570 |

**Table 10.** Thermodynamic derived properties from the first acoustic virial coefficient (equation 20 and 21) at $T$ = 325 K with their corresponding expanded uncertainties ($k$=2) and comparison with the different EoS.

| $T$ = 325 K | $Z_{exp}$ | $(Z_{exp}- Z_{GERG})/Z_{exp}$ % | $(Z_{exp}-Z_{AGA})/Z_{exp}$ % | $(Z_{exp}- Z_{PR})/Z_{exp}$ % |
|---|---|---|---|---|
| $\gamma^{pg}$ | 1.29974±0.00071 | 0.11 | 0.11 | 0.11 |
| $C_v^{pg}$ / $J\cdot mol^{-}1\cdot K^{-1}$ | 27.739±0.066 | -0.50 | -0.48 | -0.50 |
| $C_p^{pg}$ / $J\cdot mol^{-}1\cdot K^{-1}$ | 36.054±0.088 | -0.38 | -0.37 | -0.39 |
| $\beta_a$ / $m3\cdot mol^{-1}$ | $(-363.6\pm8.9)\cdot10^{-7}$ | -3.3 | -8.9 | -11 |
| $\gamma_a$ / $(m3\cdot mol^{-1})^2$ | $(38.8\pm4.2)\cdot10^{-10}$ | $-1.8\cdot10^9$ | -64 | -270 |

**Figure 1.** Relative excess half-width ($\Delta g/f_{exp}$) of radial modes as a function of pressure at $T =$ 273 K for modes: $\diamond$ (0,2), $\square$ (0,3), $\triangle$ (0,4), $\times$ (0,5), $*$ (0,6).

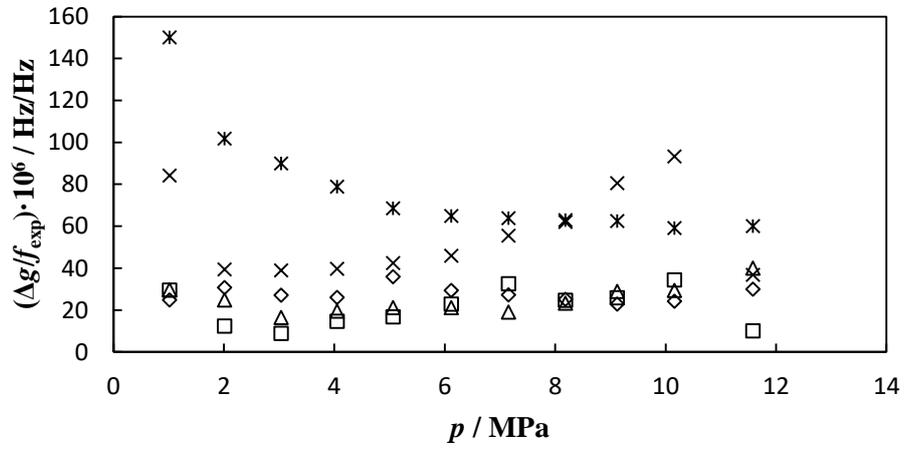

**Figure 2.** Relative excess half-width ($\Delta g/f_{exp}$) of radial modes as a function of pressure at $T =$ 300 K for modes: $\diamond$ (0,2), $\square$ (0,3), $\triangle$ (0,4), $\times$ (0,5), $*$ (0,6).

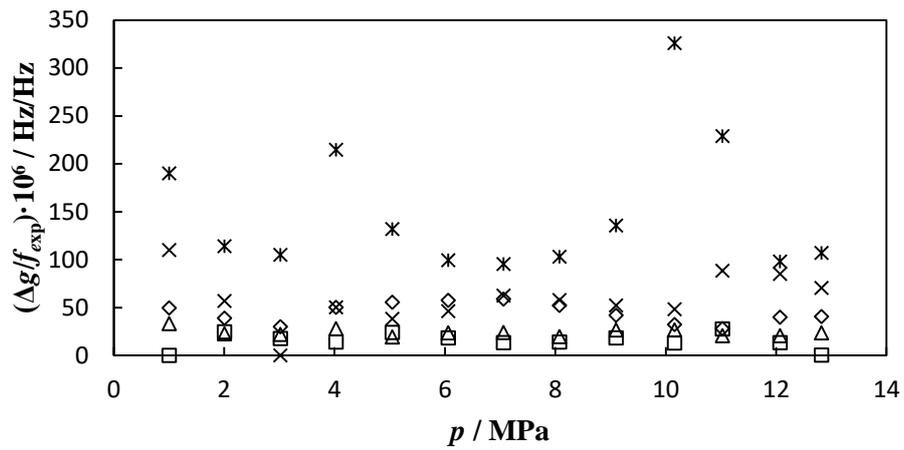

**Figure 3.** Relative excess half-width ($\Delta g/f_{\exp}$) of radial modes as a function of pressure at $T = 325$ K for modes: ◇ (0,2), □ (0,3), △ (0,4), × (0,5), ✳ (0,6).

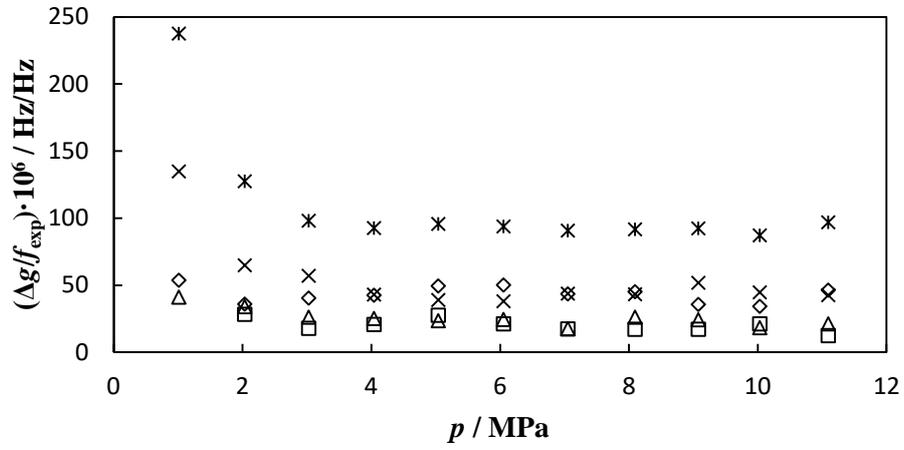

**Figure 4.** Overall relaxation constant time $\tau_{vib}$ due to vibrational relaxation of radial modes as a function of pressure at $T = 273$ K for modes: ◇ (0,2), □ (0,3), △ (0,4), × (0,5), ✳ (0,6).

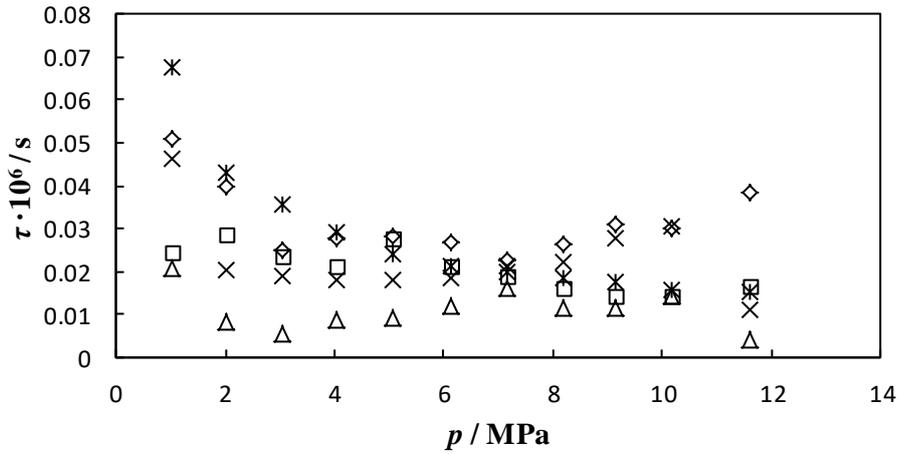

**Figure 5.** Overall relaxation constant time $\tau_{vib}$ due to vibrational relaxation of radial modes as a function of pressure at $T$ = 300 K for modes: $\diamond$ (0,2), $\square$ (0,3), $\triangle$ (0,4), $\times$ (0,5), $\ast$ (0,6).

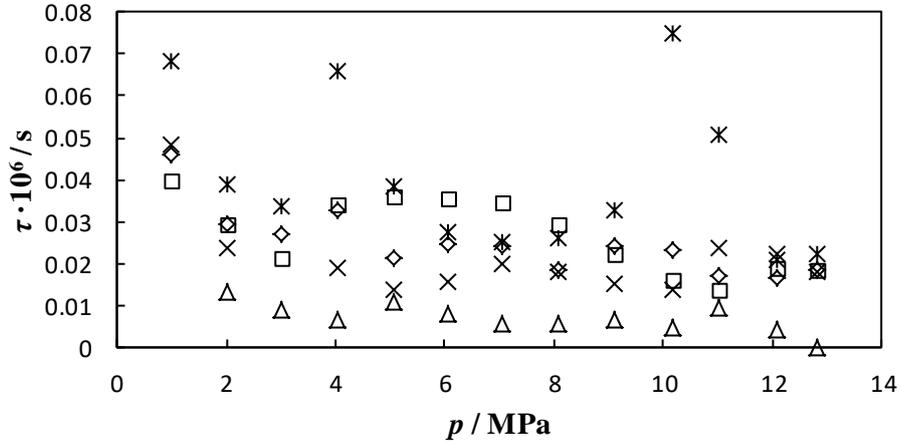

**Figure 6.** Overall relaxation constant time $\tau_{vib}$ due to vibrational relaxation of radial modes as a function of pressure at $T$ = 325 K for modes: $\diamond$ (0,2), $\square$ (0,3), $\triangle$ (0,4), $\times$ (0,5), $\ast$ (0,6).

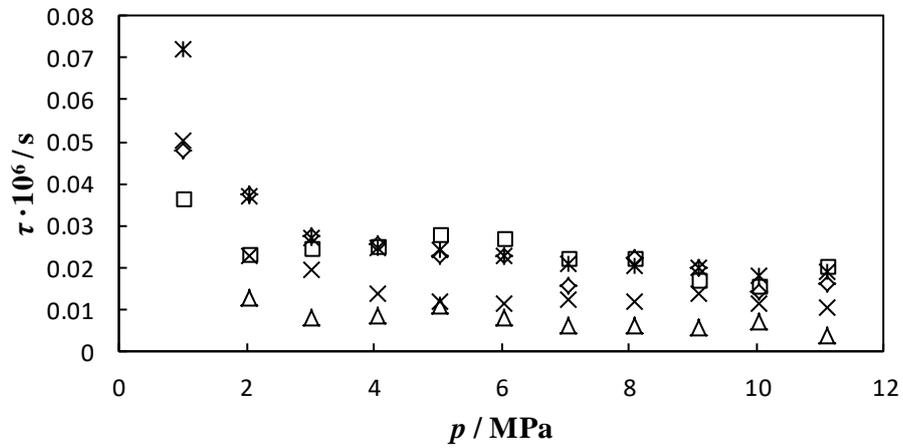

**Figure 7.** Residual analysis $\Delta c = (c_{fitted} - c_{exp})/c_{exp}$ as a function of pressure of the measured speeds of sound and the values fitted by equation 19, all within measurement uncertainty, at $T = 273$ K (×), $T = 300$ K (∗), $T = 325$ K (+).

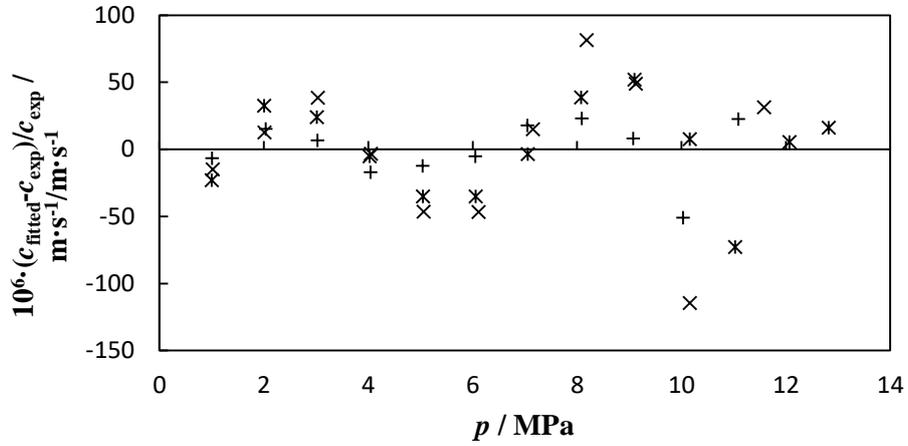

**Figure 8.** Relative deviations $\Delta c = (c_{GERG-2008} - c_{exp})/c_{exp}$ as function of pressure, expanded ($k=2$) uncertainty as a dotted line and uncertainty of model EoS GERG-2008 as a dashed line at $T = 273$ K (×), $T = 300$ K (∗), $T = 325$ K (+).

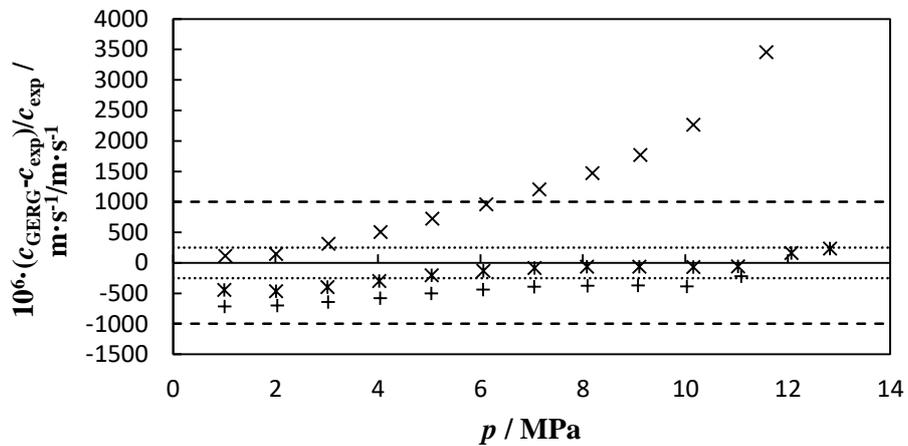